\documentclass[aps,prb,superscriptaddress,showpacs,twocolumn]{revtex4}
\usepackage{graphicx}
\usepackage{dcolumn}
\usepackage{bm}

\begin{document}

\title{Parameter Scaling in the Decoherent Quantum-Classical Transition for
chaotic rf-SQUIDs}
\author{Ting Mao}
\author{Yang Yu}
\email{ yuyang@nju.edu.cn}
\affiliation{National Laboratory of Solid State Microstructures and Department of
Physics, Nanjing University, Nanjing 210093, China }

\begin{abstract}
We numerically investigated the quantum-classical transition in rf-SQUID
systems coupled to a dissipative environment. It is found that chaos emerges
and the degree of chaos, the maximal Lyapunov exponent $\lambda _{m}$,
exhibits non-monotonic behavior as a function of the coupling strength $D$.
By measuring the proximity of quantum and classical evolution with the
uncertainty of dynamics, we show that the uncertainty is a monotonic
function of $\lambda _{m}/D$. In addition, the scaling holds in SQUID
systems to a relatively smaller $\hbar _{eff}$, suggesting the universality
for this scaling.
\end{abstract}

\maketitle

\section{INTRODUCTION}

How classical behavior arises in a quantum mechanical system is one of the
essential questions in quantum theory, and has long attracted intense
interest. The quantum to classical transition (QCT), which has been well
understood to be mainly induced by decoherence caused by the coupling with the
environment,\cite{Zurek,notes} attains some progresses in recent years. It is
proposed that the QCT is controlled by relevant parameters including the
effective Planck constant $\hbar _{eff}$ (i.e., the relative size of the
Planck constant), a measure of the coupling with the environment $D$, and
the Lyapunov exponent $\lambda $, for chaotic systems.\cite{Pattanayak} By
computing measures which directly reflect the \textquotedblleft
distance\textquotedblright\ between quantum and classical evolutions, it is
shown that the distance is controlled by a composite parameter of the form $%
\zeta =\hbar ^{\alpha }\lambda ^{\beta }D^{\gamma }$. Many efforts on
investigating the coefficients $\alpha $, $\beta $, $\gamma $ have been made
\cite{Toscano,Gammal} in different systems such as the kicked harmonic
oscillator and the Duffing oscillator. However, in the previous systems, $%
\lambda $ is generally a constant. Therefore, the direct illustration of the
effect of the Lyapunov exponent $\lambda $ on the computed distance is still
open.

In this article we try to explore the parameter scaling in QCT by using the
system of the superconducting quantum interference device (SQUID). Rf-SQUID
system has been demonstrated as a well controllable decoherent quantum
system. Macroscopic quantum phenomena such as resonant tunneling\cite{Rouse}
and level quantization\cite{Silvestrini} and quantum superposition\cite%
{Friedman} have been reported. On the other hand, the strong coupling
between the SQUID and the environment can introduce chaos. As early as 1983,
the chaotic behavior of the SQUID treated as a semi-classical model had been
found.\cite{Fesser} Recently, a research shows that a three-junction SQUID
can be used to study the dynamics of quantum chaos.\cite{Pozzo} Such works
motivate us to study the chaotic behavior of SQUID under decoherence induced
by environment, which enables us to directly demonstrate the effect of the
Lyapunov exponent on QCT.

This article is organized as follows. In Sec.II we numerically investigate
the chaotic dynamics of SQUID with coupling to an external environment, and
it is shown that the maximal Lyapunov exponent $\lambda _{m}$, which
quantifies the chaotic degree of SQUID, is non-monotonic as a function of $D$%
, a measure of the coupling. Thus we can say in some regimes of $D$, the
chaos of SQUID is suppressed by the decoherence induced by environment\cite%
{Yamazaki}. In Sec.III we use the uncertainty of dynamics as the distance
between quantum and classical evolutions, and show that the uncertainty
behaves rightly, even in the chaos suppressed region, as a monotonic
function of $\lambda _{m}/D$. To the best of our knowledge, this is the first
direct demonstration of the scaling relation since it was proposed\cite%
{Pattanayak}.

\section{chaotic dynamics of SQUID}

The rf-SQUID system considered here consists of a large superconducting loop
interrupted by a single Josephson junction with a critical current $I_{c}$.
Under the driving of a external flux $\phi _{ex}(t)$ with the form of $\phi
_{ex}(0)\cos (\omega _{d}t)$ (where $\phi _{ex}(0)$ and $\omega _{d}$
respectively denote the driving amplitude and driving frequency), the
Hamiltonian for the SQUID system can be given as
\begin{equation}
\hat{H}_{D}=\frac{\hat{q}^{2}}{2C}+\frac{(\hat{\phi}-\phi _{ex}(t))^{2}}{2L}+%
\frac{I_{c}\phi _{0}}{2\pi }\cos (2\pi \hat{\phi}/\phi _{0}),
\label{Squid Hamiltonian}
\end{equation}%
where $C$ is the junction capacitance, $L$ is the rf-SQUID inductance and $%
\phi _{0}=h/2e$ denotes the superconducting flux quantum. The magnetic flux
threading the rf-SQUID $\hat{\phi}$ and the total charge on the capacitor $%
\hat{q}$ are the conjugate variables of the system with the imposed
commutation relation $[\hat{\phi},\hat{q}]=i\hbar $.

We can rewrite this Hamiltonian into a dimensionless one\cite{Everitt} as
\begin{equation}
\hat{H}_{D}=\frac{\hat{Q}^{2}}{2}+\frac{(\hat{\Phi}-\Phi _{ex}(t))^{2}}{2}+%
\frac{I_{c}}{2\omega _{0}e}\cos (\frac{2e}{\sqrt{\hbar \omega _{0}C}}\hat{%
\Phi}),  \label{reduced Squid Hamiltonian}
\end{equation}%
in which $\omega _{0}=1/\sqrt{LC}$, $\Phi _{ex}(t)=\sqrt{\frac{\omega _{0}C}{%
\hbar }}\phi _{ex}(t)$, and $\hat{Q}=\sqrt{1/\hbar \omega _{0}C}\hat{q}$, $%
\hat{\Phi}=\sqrt{\omega _{0}C/\hbar }\hat{\phi}$ satisfy the commutation
relation $[\hat{\Phi},\hat{Q}]=i$.

Since no chaos can be seen in the dynamics of isolated quantum systems,\cite%
{Habib} to study the chaotic behaviors of the SQUID system, we couple the
system to a dissipated environment in the Markovian limit. We adopt the
quantum state diffusion (QSD) \cite{Percival} approach which is widely used
in studying open quantum systems \cite{Brun,Kapulkin,Ota} to describe the
evolution of this coupled system. The QSD equation for the evolution of the
state vector $|\psi \rangle $ reads
\begin{eqnarray}
|d\psi \rangle  &=&-\frac{i}{\hbar }\hat{H}|\psi \rangle dt+\sum_{j}\Big(%
\langle \hat{L}_{j}^{\dagger }\rangle \hat{L}_{j}-\frac{1}{2}\hat{L}%
_{j}^{\dagger }\hat{L}_{j}  \nonumber  \label{QSD} \\
&&-\frac{1}{2}\langle \hat{L}_{j}^{\dagger }\rangle \langle \hat{L}%
_{j}\rangle \Big)|\psi \rangle dt+\sum_{j}(\hat{L}_{j}-\langle \hat{L}%
_{j}\rangle )|\psi \rangle d\xi _{j},
\end{eqnarray}%
where $\hat{H}$ is the system Hamiltonian and $\hat{L}_{j}$ are the Lindblad
operators representing the coupling with the environment. $d\xi _{j}$ are
independent complex differential Gaussian random variables satisfying $%
M(d\xi _{j})=M(d\xi _{i}d\xi _{j})=0$, $M(d\xi _{i}^{\ast }d\xi _{j})=\delta
_{ij}dt$ (where $M$ denotes the ensemble mean). For the SQUID system
considered here, we have $\hat{H}$ and $\hat{L}$ for Equation (\ref{QSD}) as
$\hat{H}=\hat{H}_{D}+\hat{H}_{R}$, $\hat{L}=\sqrt{D}(\hat{\Phi}+i\hat{Q})$,
where $\hat{H}_{D}$ is shown in Equation (\ref{reduced Squid Hamiltonian}), $%
\hat{H}_{R}=\frac{D}{2}(\hat{\Phi}\hat{Q}+\hat{Q}\hat{\Phi})$ \cite%
{Brun,Kapulkin} is a damping term added to recover the correct equation of
motion in the classical limit, and $D$ is the strength of the coupling with
the environment mentioned in the beginning.

\begin{figure}[tbp]
\centering
\includegraphics[width=3.4in]{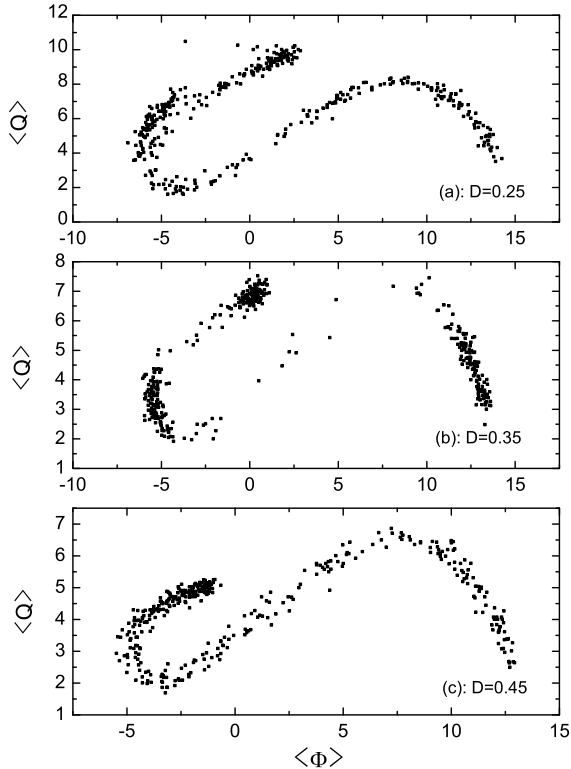}
\caption{Poincar\'{e} sections for $D=0.25,0.35,0.45,$ from top to bottom.
From middle panel we can see that points are largely confined in three
regions, which indicates a non-monotonic transition of chaos.}
\end{figure}

Using the powerful QSD library,\cite{Schack} we numerically solve the
Equation (\ref{QSD}) and investigate the change in the dynamics of the SQUID
system when increasing the strength of dissipation. A typical set of SQUID
parameters is selected here, $C=0.1pF$, $L=300pH$, $I_{c}=2.2\mu A$, $\omega
_{d}=1.14\omega _{0}$, $\phi _{ex}(0)=0.2684\phi _{0}$, which insures the
action of this system is small enough compared with fixed $\hbar $.\cite%
{Habib} Then we examine 28 different values of $D$ from slightly dissipated ($%
D=0.23$) to heavily dissipated regime ($D=1$) in our calculation, during
which we have the same initial state $|\psi (t=0)\rangle =|\sqrt{2}(\langle
\hat{\Phi}\rangle +i\langle \hat{Q}\rangle )=(0.877-0.566i)\rangle $--the
coherent state--and same realization of generating the random numbers. The
quantum Poincar\'{e} sections, which each comprises of $500$ points taken at
a fixed phase of the external driving once a driving period, are shown for
three representative values of $D$ in Fig.1(a)-1(c). It can be clearly seen
in Fig.1(a) that points forms a uniformly stretched Poincar\'{e} profile in
the phase space which indicates \textquotedblleft chaos\textquotedblright\
for $D=0.25$. However, for $D=0.35$ most of points are confined in three
relatively small regions as shown in Fig.1(b), which indicates the
suppression of chaos. Then the Poincar\'{e} profile similar to the one in
Fig.1(a) is recovered in Fig.1(c) when $D$ is increased to $0.45$. Some non-monotonic analogous phenomena have been studied in classical chaotic systems,\cite%
{Yamazaki,Matsumoto} and a qualitative explanation has been proposed there. If the chaotic attractors are narrowly and non-uniformly distributed in phase space, the fluctuation induced by dissipation may cause the neighboring trajectory jump over it, which results in the suppression of chaos. While further increasing dissipation intensity, the structure of the chaotic attractor may be modified and thus spread wider than before. Therefore the system becomes chaotic again. Since $(\langle \hat{\Phi}%
(t)\rangle ,\langle \hat{Q}(t)\rangle )$ form classical-like trajectories in our calculation, we expect that the explanation is also valid for the suppression of chaos in quantum region.
\begin{figure}[tbph]
\centering
\includegraphics[width=3.4in]{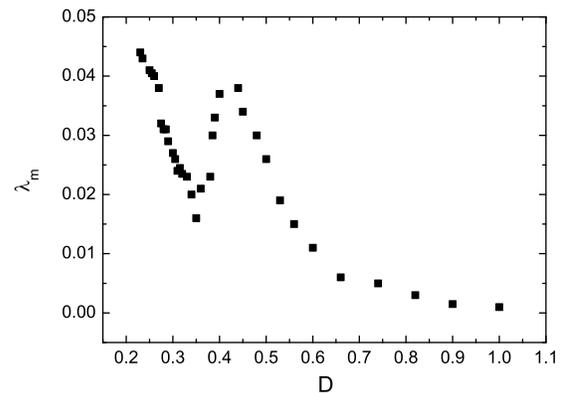}
\caption{Maximal Lyapunov exponent $\protect\lambda _{m}$ versus $D$.The
distinctive dip rightly attests the occurrence of suppression of chaos.}
\end{figure}

To describe this transition of chaos quantitatively, we calculate the
maximal Lyapunov exponent $\lambda_{m}$ for a time series--the expectation
value of the magnetic flux $\langle\hat{\Phi}(t)\rangle$--at each value of $%
D $. The calculation is based on the method and programs \cite{Kantz,Hegger}
which are specifically designed for the analysis of nonlinear time series. With carefully chosen parameters as the delay time $d=3$, the embedding
dimension $m=3$ and the scaling length $s=1.4\%$ for the calculation to best
meet the requirements in Ref.18, the sufficient convergency of the Lyapunov
exponent is guaranteed. The result is shown in Fig.2, in which the graph
of $\lambda _{m}$ versus $D$ has a distinctive dip in a approximate region
of $D=0.25\sim 0.45,$ indicating the suppression of chaos. We also repeat
the whole calculation above in some different realization of random numbers
with the SQUID parameters and the initial state fixed, and find the curves
are quite analogous to the one in Fig.2.

\begin{figure}[tbp]
\centering
\includegraphics[width=3.5in]{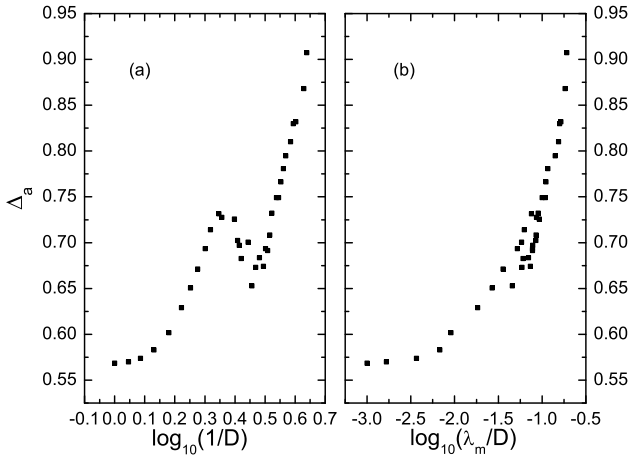}
\caption{Averaged uncertainty $\Delta _{a}$ as a function of (a) $D$ and (b)
a composite parameter $\protect\lambda _{m}/D$. The monotonic increase of $%
\Delta _{a}$ as a function of $\protect\lambda _{m}/D$ in (b) demonstrated
the scaling law.}
\end{figure}

\section{effect of maximal lyapunov exponent on QCT}

With the non-monotonic relationship between maximal Lyapunov exponent $%
\lambda _{m}$ and the strength of the coupling with the environment $D$, we
can directly investigate the effect of $\lambda _{m}$ on QCT. To measure the
\textquotedblleft distance \textquotedblright between quantum and classical evolution, we use the well known
quantity--the uncertainty of dynamics $\Delta =\sqrt{\langle (\hat{\Phi}%
-\langle \hat{\Phi}\rangle )^{2}\rangle }\sqrt{\langle (\hat{Q}-\langle \hat{%
Q}\rangle )^{2}\rangle }$, which is simple for calculation and adequate to
describe the QCT. According to the commutation relation $[\hat{\Phi},\hat{Q}%
]=i$, it follows that $\Delta \geq 0.5$. By solving Equation (%
\ref{QSD}) with same calculating parameters as in Sec.II, we get a time
series of the uncertainty $\Delta (t)$ at each value of $D$. After averaging
each series of $\Delta (t)$ over a reasonably long time ($>$ 100 periods of
the external driving), we obtained the curve of the averaged uncertainty $%
\Delta _{a}$ versus $D$ and showed in Fig.3(a), where $D$ has the same
sequence of values as in Fig.2. It can be clearly seen that in Fig.3(a) a
obvious dip emerges in the very regime where chaos is suppressed by the
dissipation, which implies QCT directly depends on the degree of chaos.
Motivated by this, we combine $\lambda _{m}$ and $D$ with the form of $%
\lambda _{m}/D$ which is inferred in Ref.2 and look into the relationship
between $\Delta _{a}$ and such composite single parameter. Shown Fig.3(b) is
an example of $\Delta _{a}$ vs. $\lambda _{m}/D$. One can find that the dip
is rubbed out and $\Delta _{a}$ approximately shows a monotonic increasing
in $\lambda _{m}/D$ with two distinct regimes of small and large increasing
rates.\cite{Pattanayak} Therefore we demonstrate the scaling between $\lambda _{m}
$ and $D$ holds over a considerable range in $\Delta _{a}$. It is noticed
that the points which lie in the dip in Fig.3(a) spread slightly around the
curve in Fig.3(b). We conjecture this spread could be mainly attributed to
the calculating error \cite{Kantz} of $\lambda _{m}$ which is induced by the
inevitable quantum noise added into the trajectory of $(\langle \hat{\Phi}%
(t)\rangle ,\langle \hat{Q}(t)\rangle )$, especially when chaos is
suppressed and the value of $\lambda _{m}$ is comparatively small.

\begin{figure}[tbp]
\centering
\includegraphics[width=3.6in]{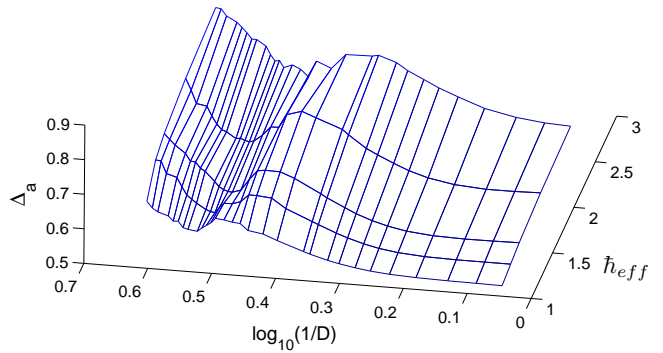}
\caption{(Color online) Averaged uncertainty $\Delta_{a}$ as a function of $D$ and $%
\hbar_{eff}$.The parameters for the system with largest effective Planck
constant $\hbar_{eff}=2.6$ has been shown in text. Other $\hbar_{eff}$ and
corresponding sets of parameters are listed in Table.I.}
\end{figure}

\begin{table}
\centering
\caption{$\hbar_{eff}$ and corresponding parameters}
\begin{tabular}{cccccc}
\hline \hline
$\hbar_{eff}$  & $I_{c}(\mu A)$  &  $L(pH)$  & $C(pF)$ & $\omega_{d}(\omega_{0})$ & $\phi_{ex}(0)(\phi_{0})$ \\
\hline
1    & 4.6 & 100& 3.95& 0.65 & 0.081\\
1.2 & 3.35 & 150& 2.16 & 0.71 & 0.1041\\
1.4 & 2.67 & 200& 1.29  & 0.78 & 0.1273\\
1.9 & 2.46 & 250& 0.36  & 0.99 & 0.1851\\
\hline \hline
\end{tabular}
\end{table}

\begin{figure}[tbp]
\centering
\includegraphics[width=3.9in]{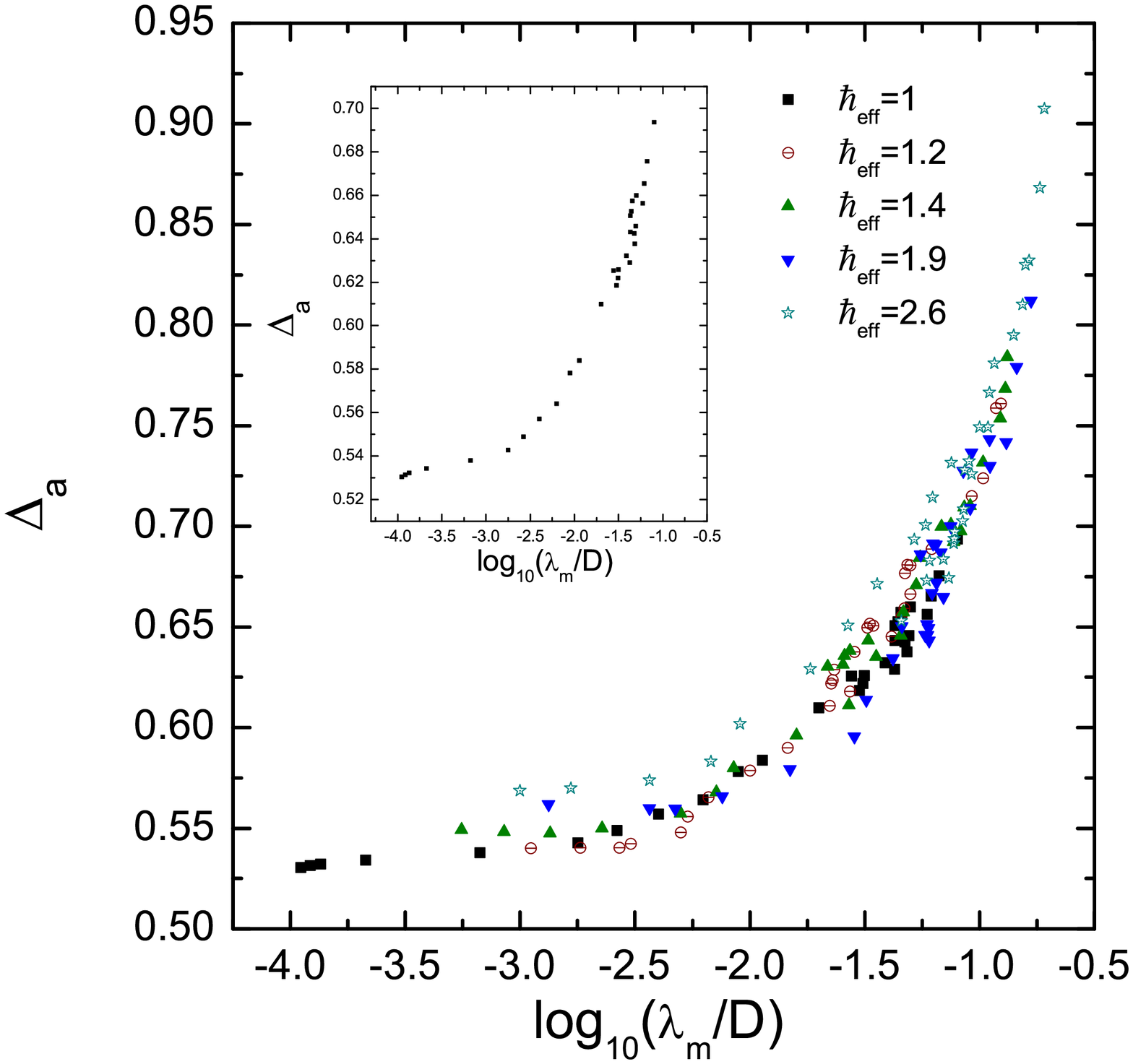}
\caption{(Color online) $\Delta _{a}$ versus a composite parameter $\protect\lambda _{m}/D$ for different effective Plank constant. It is shown that the scaling
law holds for systems with different $\hbar _{eff}$. Inset: The curve with $%
\hbar _{eff}=1$ is shown separately. }
\end{figure}

Now we examine this scaling law for the SQUID system with a smaller
effective Planck constant $\hbar_{eff}$. To obtain a smaller $\hbar_{eff}$%
, it is not straightforward for the SQUID system to directly manipulate the value of $\hbar$ .\cite{Everitt}Instead, we enlarge the action of the SQUID system simply by changing parameters in the Hamiltonian; the larger the action the smaller $\hbar_{eff}$, and vice
versa.\cite{Habib} By deliberately selecting the set of parameters including
$I_{c}$, $L$, $C$, $\omega _{d}$ and $\phi _{ex}(0)$, we can enlarge the
action and maintain the chaotic dynamics of the system at the same time. The
values of these parameters are not difficult to modulate for a realistic
SQUID system where $I_{c}$ could become controllable by replacing the
single Josephson junction with a small loop (dc SQUID) which contains two
identical Josephson junctions,\cite{Rouse} $C$ and $L$ are both under the
upper realistic limit of typical Josephson junctions. We select four sets of
parameters for the SQUID systems each of which has a smaller $\hbar _{eff}$
compared with the foregoing system's. Assuming the smallest $\hbar _{eff}$
is equal to 1 and comparing the actions of the systems which are measured
with the system size,\cite{Habib} we approximately gain the value of other
effective Planck constants as follow, 1.2, 1.4, 1.9, 2.6, where 2.6 is the
value of the foregoing system's $\hbar _{eff}$. Then we apply the same
calculating procedures to these systems, and the results are shown in Fig.4
and Fig.5 which also include the data of the foregoing system for
comparison. Fig.4 shows the averaged uncertainty $\Delta _{a}$ as a function
of $D$, $\hbar _{eff}$. For each $\hbar _{eff}$,a distinct dip exists
as expected in the region where chaos is suppressed by the dissipation of
environment. Fig.5 shows the same data plotted as a function of $\lambda
_{m}/D$, in which, the behavior of $\Delta _{a}$ for each $\hbar _{eff}$ is
considerably the same, which demonstrates the scaling between $\lambda _{m}$
and $D$ is still valid for a system with relatively small $\hbar _{eff}$.
For clarity, we separately show the curve with $\hbar _{eff}=1$ in the inset
of Fig.5. Since a larger action is helpful to undermine the effect of
quantum noise, more accurate $\lambda _{m}$ can be gained for the system
with smaller $\hbar _{eff}$, which, is reflected in the lack of noticeable
spread around the curve in the inset.

We also chose some different random numbers generator to repeat the
calculation for SQUID systems with different $\hbar _{eff}$, and succeed in
getting same qualitative conclusions as discussed above.

\section{conclusion}

In summary, we investigated QCT in chaotic rf-SQUIDs. The suppression of
chaos induced by environment dissipation was observed in quantum regime. It
is found that the quantum to classical transition in the presence of a
dissipated environment is governed by a composite parameter $\lambda _{m}/D$%
. It could be expected the scaling law between $\lambda _{m}$ and $D$ would
holds over a wide range of $\hbar _{eff}$. However, to generalize this
scaling to the one involving $\hbar _{eff}$, $\lambda _{m}$ and $D$ and to
reveal the coefficients between them are still open questions needed to
explore.

\section{ ACKNOWLEDGMENTS}

This work was partially supported by the NSFC (under Contract Nos.
10674062,10725415), the State Key Program for Basic Research
of China (under Contract Nos. 2006CB921801), and the
Doctoral Funds of the Ministry of Education of the People's Republic of
China (under Contract No. 20060284022 ).


\begin{thebibliography}{99}
\bibitem{notes} The presence of classical chaos may lead to novel quantum phenomena in some quantum systems isolated from the environment. These phenomena are out of scope of our discussion in this paper.
\end{thebibliography}
\end{document}